\documentclass[12pt,letterpaper]{article}
\usepackage{amsfonts,amsmath}
\usepackage{lscape,graphicx,epsf}

\usepackage[headings]{}

\def\ds{\displaystyle}
\def\bea{\begin{array}{c}}
\def\ea{\end{array}}
\def\be{\begin{equation}\bea\ds}
\def\ee{\ea\end{equation}}
\def\bmlt{\begin{multline}}
\def\emt{\end{multline}}

\def\d{{\rm d}}
\def\Nn{{\cal{N}}}

\def\p{{\prime}}
\def\uor{U(1)_{\cal{R}}}
\def\bt{\tilde{\beta}}
\def\At{\tilde{A}}

\newcommand\Trule{\rule{0pt}{2.6ex}}
\newcommand\Brule{\rule[-1.2ex]{0pt}{0pt}}

\begin{document}

\begin{titlepage}

\begin{flushright}
{PUPT-2243}\\
{SU-ITP-07/18}\\
{ITEP-TH-32/07}\\
{RUNHETC-2007-13}\\
\end{flushright}

\vskip1cm
\begin{center}
{\Huge\bf Gravity Multiplet \\
\vskip 0.3cm
on KS and BB backgrounds} \\
\vskip0.5cm {\Large A.~Dymarsky$^{a,b,c}$ and D.~Melnikov$^{c,d}$}
\vskip1cm
{\large\itshape $^{a}$Joseph Henry Laboratories,\\
Princeton University, Princeton, NJ 08544\\
\vspace{0.5cm}
\large\itshape $^{b}$Stanford Institute for Theoretical Physics,\\
Stanford, CA 94305\\
\vspace{0.5cm}
$^{c}$Institute for Theoretical and Experimental Physics,\\
B.~Cheremushkinskaya 25, 117259 Moscow, Russia\\
\vspace{0.5cm}
$^{d}$Department of Physics and Astronomy,\\
Rutgers, 136 Frelinghuysen Rd, Piscataway, NJ 08854\\
}
\end{center}

\vspace{0.5cm}
\begin{abstract}
In this paper we study the spectra of glueballs on the
Klebanov-Strassler background and its extension to the baryonic
branch. We numerically calculate the mass spectrum of glueballs from
the spin 2 ``gravity'' multiplet, which contains the traceless part
of the stress-energy tensor and the transverse part of the $U(1)$
${\cal{R}}$-current. The mass spectra of the corresponding
fluctuations in supergravity coincide due to supersymmetry, which
is manifest in the effective five-dimensional theory through a
Supersymmetric Quantum Mechanics transformation. We show that the
glueball spectra grow as $m_n^2\propto Un^2$ for large values
of the baryonic branch parameter $U$.
\end{abstract}

\end{titlepage}

\section{Introduction}
\label{intro} Ideas of holography underlaid the AdS/CFT
correspondence \cite{Maldacena} provide a promising perspective
for the study of gauge theories via string theory and
supergravity. In particular, the extension of the holographic
principle to a non-conformal case enables to capture the
 strongly coupled dynamics of a gauge theory through the classical
supergravity. This allows, at least in principle, to calculate
various correlation functions, extracting the masses of glueballs,
which is not possible by means of the standard field theory
technique. In this paper we focus on a particular case of the
Klebanov-Strassler supergravity background \cite{KS} and its
extension to the baryonic branch \cite{GHK,Butti,DKS}, which is a
gravity dual of the non-conformal $\Nn=1$ gauge theory with two
pairs of matter multiplets.

According to the AdS/CFT correspondence, gauge theory operators
correspond to fluctuations of the background supergravity
fields. Thus the stress-energy tensor corresponds to the
fluctuation of the metric. The transverse traceless part of the
former combines with the transverse part of the $\uor$ current
$J_5^\mu$ and the transverse fermionic superconformal current into
a spin 2 massive supermultiplet \cite{FZ}. The mass spectra of the
corresponding glueballs coincide, what is evident from the
supersymmetric structure of the equations of motion in the gravity
dual theory.

In the gauge theory the supersymmetric structure of the spin 2
multiplet is transparent through the on-shell equation of
current conservation \cite{FZ}. Using superfield notations, one has
\be
\label{superfield S}
D^{\alpha} V_{\alpha\dot{\alpha}}=\bar{D}_{\dot{\alpha}} \bar{S},
\ee
where $V_{\alpha\dot{\alpha}}=
\sigma^\mu_{\alpha\dot{\alpha}}V_\mu$  is a real supercurrent
that contains the $T^{\mu\nu}$ and the $J_5^\mu$ current,
\be
\label{supercurrent}
V^\mu={{J}}^\mu_5 - \frac{i}2\,\theta\sigma^\nu \bar{\theta}\,
T_{\nu}^{~\mu}+ {i}\,{\theta}^2\,\partial^{\mu}\bar{s} -
{i}\,\bar{\theta}^2\,\partial^\mu s  +
\frac14\,\bar{\theta}^2\theta^2
\left(2D^\mu+\Box{{J}}_5^\mu\right)+{\rm fermions}.
\ee
The supercurrent $V^\mu$ contains two supermultiplets: the
transverse spin-2 multiplet, which consists of the traceless
transverse components of $T^{\mu\nu}$ and the transverse part of
$J_5^\mu$, and the chiral multiplet $S$ containing the trace
$T_\mu^{~\mu}$, the divergence $\partial_\mu J_5^\mu$ and the
$\gamma$-trace of the superconformal current. It also contains the
complex scalar field $s$ as its lowest component. The chiral
multiplet $S$ thus accounts for the anomalies of the scale, $\uor$
and the superconformal symmetries, associated with the components
of the supercurrent, while the equation (\ref{superfield S}) is
the supersymmetric generalization of the anomalous divergence of
the current.

On the gravity dual side, background fluctuations of supergravity
fields that are dual to the operators in (\ref{supercurrent}) are
massless in the five-dimensional sense if the theory is conformal. If the conformal symmetry
is broken, the dual of the $\uor$ current satisfies the equation
for a massive vector particle in five dimensions. Note that the
dual of the traceless transverse part of the stress-energy tensor is
described by a five-dimensional massless equations in both
cases. However the finite warp-factor at the tip of the conifold
in the non-conformal case leads to a finite four-dimensional
spectrum of glueballs.

 In this work we describe
the holographic dual modes of the traceless part of
(\ref{supercurrent}), namely the spin 2 \emph{gravity} multiplet,
in the context of the baryonic branch of the KS background. The
baryonic branch is a continuous family of the type IIB
supergravity solutions originating at the KS background. The
branch is parameterized by the vevs of the baryonic operators in
the dual gauge theory \cite{GHK, BDK}. The backgrounds from the
family are constructed in terms of the Papadopoulos and Tseytlin
\cite{PT} ansatz, which consists of scalar functions
parameterizing the metric and fluxes. Those scalar functions depend
only on the radial coordinate of the conifold $t$ and satisfy a
system of first order  differential equations, which was derived in
\cite{Butti}. No analytical solution to this system is known,
except for the KS case\footnote{The analytical solution \cite{MN},
known as Maldacena-Nunez background, also solves the system of
\cite{Butti}. However, it has different boundary conditions at
infinity and therefore does not belong to the baryonic branch.},
and in practice the backgrounds from the baryonic branch have to
be constructed numerically. More details about the baryonic branch
solutions can be found in the works \cite{Butti,DKS}.

The KS and baryonic branch backgrounds correspond to
non-conformal gauge theories. We derive the linearized equations
for the vector fluctuation dual to the $U(1)$ $\cal{R}$-current
and the metric fluctuation dual to the stress-energy tensor. Then
we numerically compute the four-dimensional mass spectra of the
corresponding glueballs along the branch. Since we derive our
equations on the solid ground of ``microscopic'' ten-dimensional
theory, we can test the applicability of the similar results
obtained in \cite{DeWolfe} through an effective approach of
five-dimensional models of gauge/gravity correspondence.

This paper is
organized as follows.  In the section \ref{holographic anomaly} we
remind the reader the dual description of the bosonic operators of
the gravity and anomaly multiplets in the case of the KS
background. This part also contains a sketch of the holographic
anomaly mechanism suggested in \cite{Ouyang}.

Sections \ref{gm section} and \ref{subsection vector} are
dedicated to a derivation of the linearized equations for the
bosonic fluctuations of fields dual to the gravity multiplet of (\ref{supercurrent}). On the supergravity side the bosonic sector of the
multiplet consists of the symmetric traceless
perturbation of the metric -- the graviton, and the transverse
vector perturbation discussed in the section~\ref{holographic
anomaly}.

Since the bosonic fluctuations of the gravity multiplet are related by
supersymmetry, their spectra coincide. There is a Supersymmetric
Quantum Mechanics (SQM) transformation relating the effective
five-dimensional equations, which is a reminiscence of the
original supergravity transformation in ten dimensions. In section
\ref{subsection vector} we derive the equation for the vector mode
only for the case of the KS background, but the supersymmetric structure of
the equations allows us to extend it further to the baryonic branch.
We show that this equation is the same as discovered by
\cite{DeWolfe} in the five-dimensional approach. This is discussed
in detail in section \ref{ss supersymmetry}.

We present the results of a numerical calculation of the
spectrum in the section \ref{numerics}. Although the equations
that describe the graviton and the vector particle yield the same
spectrum of bound states, they are essentially different. We
perform two separate calculations of the spectrum of the gravity
multiplet which is an important consistency check of the numerical
results. We conclude with a discussion in section
\ref{conclusions}.

\section{Multiplets and Anomalies in the Dual Theory}
\label{holographic anomaly}
The purpose of this work is the study
of the gravity multiplet, i.e. the fluctuations above a
classical supergravity background dual to a field theory
supermultiplet  consisting of the traceless part of the
stress-energy tensor $T^{\mu\nu}$, spin 3/2 conformal supercurrent
and the conserved part of the $\uor$ current $J^\mu_5$. The
gravity multiplet therefore contains the traceless symmetric
excitation of the metric -- the graviton $h_{\mu\nu}$, the spin
3/2 gravitino and the transverse vector excitation $\At_\mu$ along
the 1-form $\d\psi$ which we specify below. Classical supergravity backgrounds that we study in this work are the backgrounds from
the baryonic branch of the KS solution.

First we consider the transverse non-diagonal fluctuation of the
background metric
\be
\label{def graviton}
g_{\mu\nu}=
\eta_{\mu\nu} + h_{\mu\nu},
\ee
where $h_{\mu\nu}$ has only
components in the Minkowskian directions. We find that in
accordance with the general results of \cite{GKT,Tye} this
excitation is described by the massless scalar minimally coupled
to the metric in the Einstein frame for all backgrounds of the baryonic branch.

Next we consider  fluctuation that is dual to the conserved
(transverse) part of the  $\uor$ current in the KS background. Recall that the
KS solution corresponds to a manifold that is locally a product
of the Minkowski space-time and the six dimensional deformed
conifold \cite{KS},
\be
\label{10d metric} \d s_{\rm KS}^2 =
h^{-1/2}(t)\eta_{\mu\nu}\d x^\mu\d x^\nu + h^{1/2}(t)\d s_6^2,
\ee
where $h(t)$ is the warp factor that depends on the radial
coordinate $t$ of the conifold, related to the standard
conical radial coordinate $r$ via $t~\sim~3\log r$. $\uor$
transformations act as rotations along the conifold base
$T^{1,1}$,
$$\psi\rightarrow \psi + \zeta,$$
where $\psi$ is one of the angles on  the base.\footnote{One can think
of $T^{1,1}$ as of the space $S^3\times S^3/U(1)\simeq S^3\times
S^2$. The angle $\psi$ is obtained by the identification of the
3rd Euler angles of both $S^3$.}

In the conformal case, the background is invariant under this
symmetry, what results in a massless gauge field $\At_\mu$. The KS
background, as well as the backgrounds along the baryonic branch,
breaks the $\uor$ symmetry already in the UV. The 2-form potential
for the RR form $F_3$ has an explicit $\psi$ dependence. In the
UV limit
$$C_2\simeq M\psi\,\omega_2,$$
where $M$ is the flux of $F_3$  through the $S^3$ of $T^{1,1}$,
and $\omega_2$ is the $\psi$ independent 2-form on $T^{1,1}$.
Given that $\psi$ itself is a double cover of the circle, $C_2$
breaks $\uor$ down to $\mathbb{Z}_{2M}$ in the UV. In the IR the
metric has an explicit $\psi$ dependence that breaks
$\mathbb{Z}_{2M}$ further to $\mathbb{Z}_2$ in the full agreement
with the gauge theory.

As a result, the corresponding fluctuation of the background
acquires mass that is not vanishing even in the UV region
\cite{Ouyang,Krasnitz thesis}. The fluctuation in question
modifies the metric along the $\psi$ direction $g_{\mu\psi}$ and
can be described by the perturbation of the 1-form $\d\psi$ by the
``gauge'' field $\At =\At_\mu\, \d x^\mu+\At_t\, \d t$,
\be
\label{vector fluctuation}
\d\psi\rightarrow \d\psi + \At .
\ee

Since the dependence on the angles of the conifold is not
important, we can restrict our attention to the five-dimensional
theory. In the conformal case, in the absence of the 3-form fluxes,
the five-dimensional vector field $\At$ satisfies the equation for
the massless  vector
\be
\label{massless vector} \d *_5 \d\At=0.
\ee
The longitudinal part of $\At$ is not fixed by the equation
(\ref{massless vector}) as it is a gauge degree of freedom.
The corresponding symmetry is anomaly free. After adding
the fluxes, the equation for $\At$ can be brought to the form
\be
\label{massive vector}
\d(f*_5\d (g\At))+ *_5\At=0,
\ee
with some
background-dependent functions $f$ and $g$. The longitudinal part
of $\At$ is no longer trivial and satisfies
\be
\label{massive longitudinal}
\d*_5 \At = 0\ .
\ee
For an observer in four
dimensions, the five-dimensional no-source equation (\ref{massive
longitudinal}) is precisely the equation with an anomalous source
\be
\label{anomaly}
\partial^\mu \At_\mu = \theta(\Lambda)\ ,
\ee
where $\mu$ denotes the space-time indices. This holographic
anomaly mechanism is discussed in more detail in \cite{Ouyang}.

The backgrounds we are interested in have a global $SU(2)\times
SU(2)$ symmetry. Since we are interested in the uncharged sector,
all fluctuations should be $s$-waves with respect to the
directions along the base of the conifold. This is obvious for the
four-dimensional metric fluctuations as we keep it
angle-independent. In the case of the vector, it is more tricky. In
fact we need to switch from the 1-form $\d\psi$ to the invariant
extension $g^5\rightarrow g^5+\At$, where
\be
\label{g5}
g^5=\d\psi + \cos\theta_1\d\phi_1 + \cos\theta_2\d\phi_2\ .
\ee
Apparently the shift of $\d\psi$ results in the same shift of
$g^5$.

The anomalies of the scale, superconformal and $\uor$ symmetries
form a chiral supermultiplet \cite{FZ}. Its bosonic part on the
gravity side contains the fluctuations of the metric trace
$h_\mu^{~\mu}$ and the longitudinal part of the vector field
$\At_\mu=\partial_\mu \tilde{a}$.

In the section \ref{subsection vector} we derive the equation
(\ref{massive vector}) for the transverse part of the vector
fluctuation (\ref{vector fluctuation}). The transverse component
decouples from the longitudinal part and from other supergravity
fluctuations. Unfortunately it is much more complicated to derive
the equation for the longitudinal mode $\tilde{a}$. Moreover there are
certain indications that it does not decouple from the other
fields and needs to be considered as a part of a more complicated
system \cite{BHM1}.  Coupling with different supergravity
excitations will lead to some non-trivial right hand side of the
equation (\ref{anomaly}). It is particularly interesting to find
the supergravity expression for $\theta(\Lambda)$ and compare it
with the gauge theory predictions. This task is more ambitious and
we leave it for a future work.

\section{Graviton Equations}
\label{gm section}
In the current and the following sections we will be interested in the
equations for the bosonic components of the gravity multiplet, the
graviton $h_{\mu\nu}$ and the vector mode $\At_\mu$. We  start
with a ten-dimensional analysis of the linearized supergravity
equations for the graviton excitations, valid for any solution on
the baryonic branch, and proceed with a derivation of the equations
for the vector field in the KS background in the section
\ref{subsection vector}.

The traceless
symmetric perturbation of the metric is described by the
five-dimensional Klein-Gordon equation for a minimal scalar
coupled to the background in the Einstein frame. A
straightforward check \cite{DM1} shows that this property holds for the whole
baryonic branch.

Here we follow the notations of Papadopoulos and Tseytlin
\cite{PT}. The functions $A(t)$, $p(t)$, $x(t)$ and $\Phi(t)$
below are scalar functions from the PT ansatz depending on the
radial variable $t$. In particular, $A(t)$ is equivalent to the warp factor in the KS case $e^{-2A}=h^{1/2}$. It should not be confused with the vector fluctuation of the metric $\At=\At_i\,\d x^i$. In the Einstein frame the equation for the
fluctuation of the graviton $\delta \left(\d
s^2\right)=e^{-2A}h_{\mu\nu}\,\d x^\mu \d x^\nu$ takes the form
\begin{equation}
\label{minimal scalar PT}
\ddot{h}_{\mu\nu}+ 2(\dot{x}-\dot{\Phi}+2\dot{A})\dot{h}_{\mu\nu} - k^2e^{-2A-6p-x} h_{\mu\nu}=0,
\end{equation}
where $k^2$ is the square of the 4-momentum and the over dots stand
for $t$ derivatives. The equation (\ref{minimal scalar PT}) is
precisely the Klein-Gordon equation for the minimal scalar in the
the baryonic branch backgrounds including the KS point.

To proceed to the explicit form of the equation (\ref{minimal scalar PT}) for the KS background one chooses
\be
\label{KS case}
e^{-2A}=h^{1/2},\qquad \ds e^{6p+2x}=\frac32\left(\coth t-t\,{\rm csch}^2 t\right),
\\ \ds e^\Phi=e^{\Phi_0}=1, \qquad e^{2x}=\frac{1}{16}\left(\sinh t\cosh t - t\right)^{2/3}h,
\ee
where $h(t)$ is the warp factor of the metric (\ref{10d metric}):
\be
\label{h}
h(t)=\int\limits_t^\infty {\rm d}x\frac{(x\coth x-1) (\sinh2x
- 2x)^{1/3}}{\sinh^2x}\, .
\ee

With these assignments the equation takes the familiar form
\cite{Krasnitz0}
\be
\label{minimal scalar KS}
\ddot{h}_{\mu\nu}\,+\frac83\frac{\sinh^2 t}{\sinh2t - 2t}
\,\dot{h}_{\mu\nu}\, -  {k^2}\, \frac{h(t)\sinh^2t} {(\sinh 2t -
2t)^{2/3}}\,h_{\mu\nu}=0.
\ee
In the last term  we absorbed the
numerical constants in the normalization of the
momentum. It is also convenient to write the equation in the
conventional Shroedinger form
\be
\label{Shroedinger form}
(-\partial^2_t+V_2(t))h_{\mu\nu}=0,
\ee
with the effective
potential $V_2(k^2,t)$ given by
\be
\label{graviton potential}
V_2=\frac{k^2\,h(t)\sinh^2 t}{(\sinh 2t -
2t)^{2/3}}-\frac{8}{9}\,\frac{\sinh^4t}{(\sinh t\cosh t - t)^{2}}\,
+ \frac{4}{3}\,\frac{\sinh t\cosh t}{(\sinh t\cosh t -
t)}\,.
\ee

\section{Vector Mode}
\label{subsection vector}
To find a supergravity excitation that
corresponds to the $\uor$ current $J^{\,\mu}_5$, one should
consider a special deformation along the angular direction
${\partial/\partial \psi}$ of the $T^{1,1}\simeq S^3\times S^2$, as
was discussed in the section \ref{holographic anomaly}. We perturb
the $SU(2)\times SU(2)$ invariant 1-form $g^5$ (\ref{g5}) in the
following way:
\be
\label{delta g5}
g^5\rightarrow g^5+ 2\bt(t)\At, \qquad \At\equiv\At_\mu \d x^\mu\ ,
\ee
where $\At$ is
a 1-form describing the vector mode and $\bt(t)$ is yet unknown
function of $t$. Such a deformation leads to the following
perturbation of the metric:
\be
\label{deltag}
\d s^2\rightarrow \d s^2+ 2\,l(t)\, g^5\cdot \At,
\ee
where we introduced $l=2\bt e^{-6p-x}$ for a latter convenience. This change of the metric
will affect the Einstein equation as well as other equations of
the type IIB supergravity. In particular one needs to modify the
RR 5-form $F_5$ to preserve its self-duality:
\bmlt
\label{deltaF5}
\delta F_5 = -\beta \At\wedge \d g^5\wedge \d g^5 + \beta \d\At\wedge g^5\wedge \d g^5  + \beta e^{3p+x/2}*_5\d\At\wedge \d g^5 + \\
 +2e^{-2x}(\beta - \bt K)e^{-3p-x/2}*_5\At\wedge g^5.
\end{multline}
Here $\beta(t)$ is yet another function  to be determined and
$K=4\dot{A}\,e^{2x}$ in the PT notations \cite{PT}. This turns out to be a
minimal ansatz required for the KS solution. One can show that
there is no need to perturb the other type IIB fields if one is
interested only in the four-dimensional transverse part of $\At$.

The ansatz so far contains the unknown functions $\beta$,  and
$\bt$ or $l$, which can be fixed by the equations of motion. The
Bianchi identity provides us with the following equations:
\be
\label{vector Bianchi}
\d(\beta e^{3p+x/2}*_5\d\At) +
2e^{-2x}(\beta - \bt K)e^{-3p-x/2}*_5\At=0, \ee and a simple
equation for the function $\beta$,
\be \label{ansatz solution 0}
\dot{\beta}=0, \qquad\mbox{or}\qquad \beta=\beta_0.
\ee

To find the function  $\bt(t)$, or $l(t)$, one should linearize the
Einstein equation with the perturbation of the metric as in
(\ref{deltag}). The only nontrivial equation comes from the
$\delta R_{\mu\psi}$ term. After certain simplifications one can
write it in the form
\begin{multline}
\label{vector Einstein}
\partial^2_t \At_\mu + \left(2({\dot{l}}/{l})+6\dot{p}+3\dot{x}+2\dot{A}\right)\partial_t \At_\mu - k^2e^{-2A-6p-x} \At_\mu +
\\ +\left(   ({\ddot{l}}/{l})+({\dot{l}}/{l})(6\dot{p}+3\dot{x}+2\dot{A})-2\dot{A}(6\dot{p}+\dot{x}
)-2e^{-12p-4x}\right)\At_\mu = \\  = \left(\frac{ e^{-6p-x}}{24}\left(H_3^2+F_3^2\right) - \frac{2\beta_0}{l}\,{e^{-6p-5x}}
 K+\frac12\,e^{-4x}
  K^2\right)\At_\mu.
\end{multline}
In the KS background the square of the 3-forms is given by
\be
F_3^2=H_3^2=3e^{6p-x }\,\frac { t^2+2\,t^2
 \cosh^2t -6t\,\sinh t\cosh t + \cosh^2t - 2+ \cosh^4t}{ \sinh^4t}\,.
\ee
 If one now writes the equation (\ref{vector
Bianchi}) in components, taking into account (\ref{ansatz solution
0}) and the transversality condition  $\partial^\mu\At_\mu=0$,
\begin{multline}
\label{eqn1 ks}
\partial^2_t \At_\mu + (6\dot{p}+\dot{x}+2\dot{A})\partial_t\At_\mu - k^2e^{-2A-6p-x}\At_\mu+ \\ +\left(8\bt{\dot{A}}\,e^{-12p-2x}/\beta_0-2\,e^{-12p-4x}\right)\At_\mu =0,
\end{multline}
and compares it with the equation (\ref{vector Einstein}), one will find that two equations coincide only for
\be
\label{ansatz solution}
\beta_0=1,\qquad \mbox{and}\qquad l=e^{-x}.
\ee

Thus, the equation (\ref{eqn1 ks}) with the solution (\ref{ansatz
solution}) describes the transverse  vector excitation of the KS
supergravity solution. For computation of the mass spectrum  it is
worth writing
(\ref{eqn1 ks}) in terms of the explicit solution (\ref{KS case}).
We obtain the equation
\be
\label{vector ks}
\partial^2_t \At_\mu + {\cal{P}}(t)\,\partial_t \At_\mu +{\cal{Q}}(t)\,\At_\mu =0,
\ee
with\footnote{Here we use the same momentum normalization as in the equation (\ref{minimal scalar KS}).}
\be
{\cal{P}}(t)=\frac43\,\frac{\sinh^2 t}{(\sinh t\cosh t-t)}-2\coth t - \frac{\dot{h}}{h}\,,
\ee
\begin{multline}
{\cal{Q}}(t)= -\,{\frac {k^2h \sinh^2 t}{ \left( \sinh  2t  - 2t \right) ^{2/3}}}
-  \frac{8}{9}\,\frac{\sinh^4t}{(\sinh t\cosh t-t)^{2}} ~  - \frac{2}{3}\, \frac{\dot{h}\sinh^2t}{(\sinh t\cosh t-t)h}\,.
\end{multline}

Again, one could write the above equation in the form
(\ref{Shroedinger form}) with the new effective potential $V_1(k^2,t)$,
\begin{multline}
\label{vector potential}
V_1=\frac12\,\dot{\cal{P}} + \frac14\,{\cal{P}}^2 - {\cal{Q}}=\frac{{k^2}h\,\sinh^2t}{(\sinh 2t - 2t)^{2/3}} -1+2\coth^2t +\frac{1}{4}\,\frac{(\sinh 2t- 2t)^{4/3}}{h\sinh^4 t}+
\\  +\frac{3}{4}\frac{(\sinh 2t- 2t)^{2/3}(t\coth t -1)^2}{h^2\sinh^4t}  +\frac{2}{3}\,\frac{t\coth t-1}{(\sinh 2t- 2t)^{2/3}h}-
\\  -\frac{2(\sinh 2t- 2t)^{1/3}(t\coth t -1)\coth t}{h\sinh^2 t} .
\end{multline}

Closing this section we notice that the equation (\ref{vector ks}) presented here coincides with the equation derived by Krasnitz in the UV limit of the KS theory. The $t\rightarrow\infty$ limit of (\ref{vector ks}) is the same as the equation (4.30) of \cite{Krasnitz thesis} with the assignment
$$
W_\mu = - \frac{27}{hr^4}\,K_\mu,
$$
and the change to the standard radial variable $r=e^{t/3}$.

\section{Supersymmetry and 5d Approach}
\label{ss supersymmetry}

In this section we compare our findings with the results obtained
in the effective five-dimensional models of gauge/gravity
correspondence \cite{DeWolfe} and show that the equations for the
graviton and the vector mode are related by a Supersymmetric
Quantum Mechanics transformation.  This allows us to extend the
equation for the vector mode to the baryonic branch.

The authors of \cite{DeWolfe} systematically  study the
${\cal{R}}$-symmetry invariant sector of fluctuations above the
 $\Nn=2$ backgrounds of the five-dimensional $\Nn=8$ gauged
supergravity. Those also include the gravity multplet, i.e. the
traceless four-dimensional metric fluctuation and the vector
fluctuation, dual to the $\uor$ current.

Although the KS solution truncated to five dimensions would
correspond to a more general $\Nn=2$ supergravity theory
\cite{Ceresole}, it is nevertheless interesting to compare the
results of the two approaches. In fact, in both cases, the unbroken
supersymmetry is $\Nn=2$ as we deal with the supergravity dual models of
$\Nn=1$ gauge theories. Therefore the results based on the
on-shell supersymmetry can be applicable in both cases. Indeed, we
find that SQM transformations that relate the equations for
the graviton and the vector mode in the case of the KS background
coincide with the supergravity transformations used in
\cite{DeWolfe}.

In five-dimensional theories one can use the gauge freedom to
recast the background metric into the \emph{kink} form
\be
\label{5d metric}
\d s_5^2=\d q^2\, + e^{2T(q)}\eta_{\mu\nu}\d x^\mu \d x^\nu.
\ee

According to a general observation of \cite{GKT}, the traceless
graviton fluctuation $h_{\mu\nu}$ in five dimensions satisfies the
equation for a scalar minimally coupled to the geometry (\ref{5d
metric}),
\be
\label{ms 5d}
\left(\partial_q^2 + 4T^\p\,\partial_q
- e^{-2T}k^2 \right)\, h_{\mu\nu} =0.
\ee
Using the
transformations of the effective $\Nn=2$ supergravity of
\cite{DeWolfe} one can transform the graviton $h_{\mu\nu}$ into
its superpartner -- vector field $\hat{B}_\mu$. As a result, the
minimal scalar equation transforms into
\be
\label{vector DeWolfe}
\left(\partial_q\, e^{2T} \partial_q - k^2 + 2e^{2T}\,\frac{\partial^2 T }{ \partial
q^2}\right)\hat{B}_\mu=0.
\ee
Here again $k$ is a 4-momentum. We are
going to show that $\At_\mu$ of (\ref{vector ks}) and
$\hat{B}_\mu$ are related by a simple field redefinition.

The approach of \cite{DeWolfe} uses the superpotential, what can
be problematic for the backgrounds from the baryonic branch
(except for the KS solution) since the corresponding superpotentials
are not known. Therefore there is a concern that the equations
obtained for the KS may not be applicable for the outer branch.
Nevertheless, we notice that the equation itself is
$W$-independent. This already suggests that it is actually
valid for any background of the form (\ref{5d metric}). Below we
will give an argument based on supersymmetry that the equation
(\ref{vector DeWolfe}) can be applied to the whole baryonic branch.

Let us first show that
the equation (\ref{vector DeWolfe}) is the same as the equation
(\ref{vector ks}) after an appropriate field redefinition. One can think  of the metric (\ref{5d metric}) as an effective
metric obtained by truncation of the ten dimensional theory with
the metric (\ref{10d metric}) in the PT form, taken in the Einstein
frame,
\be
\label{10d metric PT} \d s_{10}^2= \left(e^{-6p-x}\d
t^2 + e^{2A}\eta_{\mu\nu}\d x^\mu \d x^\nu +
g^{(5)}_{\alpha\beta}\d y^\alpha \d y^\beta\right)e^{-\Phi/2}.
\ee
The metric (\ref{5d metric}) is then
\begin{multline}
ds_5^2 = \left(e^{-6p-x}\d t^2 + e^{2A}\eta_{\mu\nu}\d x^\mu \d
x^\nu\right){\det}^{1/3}(g^{(5)})e^{-4\Phi/3}=\\ \left(e^{-6p-x}\d
t^2 + e^{2A}\eta_{\mu\nu}\d x^\mu \d
x^\nu\right)e^{-2p+x}e^{-4\Phi/3},
\end{multline}
what gives the following identification for the coordinate $q$
and the function $T(q)$:
\be
\label{T and q}
\frac{\d}{\d q} \, =
\, e^{4p+2\Phi/3}\frac{\d}{\d t}, \qquad 2T=2A - 2p + x-\frac43\,\Phi.
\ee
Hence the equations for the graviton in ten and five
dimensions coincide, because they are just minimal scalar equations.

The equation (\ref{vector DeWolfe}) in the PT notations takes the
form
\begin{multline}
\label{eqn2 DeWolfe}
\partial^2_t \hat{B}_\mu + (2\dot{p}+\dot{x}+2\dot{A} -\frac23\,\dot{\Phi})\,\partial_t \hat{B}_\mu - k^2e^{-2A-6p-x}\hat{B}_\mu + \\ +\left((4\dot{p}+\frac23\,\dot{\Phi})\,(2\dot{A}-2\dot{p}+\dot{x}-\frac43\,\dot{\Phi})+ 2\ddot{A}-2\ddot{p}+\ddot{x}-\frac43\,\ddot{\Phi}\right)\hat{B}_\mu =0.
\end{multline}
To compare this to (\ref{vector ks}), derived in KS, set $\Phi=0$. To match the kinetic terms in
two equations one should redefine the field
$\hat{B}_\mu=e^{2p}\At_\mu$. After redefinition one gets
\begin{multline}
\label{eqn2 ks}
\partial^2_t \At_\mu + (6\dot{p}+\dot{x}+2\dot{A})\,\partial_t\At_\mu - k^2e^{-2A-6p-x}\At_\mu\, +\left(2\dot{p}\,(6\dot{A} + 3\dot{x})+ 2\ddot{A}+\ddot{x}\right)\At_\mu =0,
\end{multline}
which is precisely the equation (\ref{vector ks}) for the KS solution (\ref{KS case}).

We can further reduce the five-dimensional equations (\ref{ms 5d})
and (\ref{vector DeWolfe}) to one dimension by taking the square
of momentum $k^2$ to be the eigenvalue $-m^2$. This will reduce
the supersymmetry algebra to the Supersymmetric Quantum Mechanics
with two differential operators $Q_1$ and $Q_2$ that relate
the solutions of the two equations (\ref{ms 5d}) and (\ref{vector
DeWolfe}). These operators realize the effective transformations
of the supersymmetry algebra that was studied in \cite{DeWolfe}.
Indeed, there are operators $Q_1$ and $Q_2$, such that the
equations
\be
\label{SQM equations}
Q_1Q_2h_{\mu\nu}=-m^2 h_{\mu\nu} \qquad \mbox{and} \qquad
Q_2Q_1\hat{B}_\mu=-m^2\hat{B}_\mu
\ee
coincide with the equation
for the graviton (\ref{minimal scalar KS}) and the equation for
the vector mode (\ref{vector ks}) in the form (\ref{eqn2
DeWolfe}). It is easy to show that the operators that satisfy
(\ref{SQM equations}) are
\be
\label{gm sqm}
Q_1=(\partial_q+2T')=e^{4p+2\Phi/3}\left(\partial_t+2\dot{A}-2\dot{p}+\dot{x}-\frac43\,\dot{\Phi}\right)
\ee
and
\be
\label{gm sqm2}
Q_2=e^{2T}\partial_q=e^{2A+2p+x-2\Phi/3}\partial_t.
\ee
The operator $Q_2$ is precisely the operator from (73) of
\cite{DeWolfe} that realizes an $\Nn=2$ supergravity
transformation relating $h_{\mu\nu}$ and $\hat{B}_\mu$.

To get a more conventional representation of the SQM here, one can change the coordinates to $\partial_q=e^{-T}\partial_u$ and bring the equations (\ref{ms 5d}) and (\ref{vector DeWolfe}) to the form (\ref{Shroedinger form}) by redefining the wave functions $h_{\mu\nu}$ and $\hat{B}_\mu$. Let us define an operator
\be
\label{Q}
Q=\left(
\begin{array}{cc}
0 & \partial_u - W
\\ \partial_u + W & 0
\end{array}
\right)
\ee
with $W=-3T^\prime/2$, that acts on the vector made of
redefined wave functions $\psi_h$ and $\psi_B$. According to the
equations (\ref{ms 5d}) and (\ref{vector DeWolfe}) the action of
$Q^2$ is as follows
\be
\label{hamiltonian SQM}
Q^2\,\left(
\begin{array}{c}
\psi_h
\\ \psi_B
\end{array}
\right) = -m^2\left(
\begin{array}{c}
\psi_h
\\ \psi_B
\end{array}
\right) .
\ee
Therefore $Q^2$ is analogous  to the Hamiltonian of
the SQM. Notice, however, that its eigenvalues are $m^2$, not $m$,
because $Q_1$ and $Q_2$ correspond to the squares of the original
supersymmetry transformations, i.e. $Q_1,Q_2$ are  bosonic
operators.

We see now that the equation (\ref{ms 5d}) and (\ref{vector
DeWolfe}) are related by supersymmetry transformation for any
background (\ref{5d metric}). Since the minimal scalar equation
describing the graviton is valid for the whole branch, the
superpartner of the graviton (the transverse vector mode)
satisfies the ``superpartner'' equation (\ref{vector DeWolfe}) for
any background from the baryonic branch.\footnote{ In general, there
is a family of equations like (\ref{vector DeWolfe}) that are
related to (\ref{ms 5d}) by a supersymmetry transformation. Indeed,
for a given $W$ from (\ref{Q}), any $\hat{W}$ that satisfies
$\hat{W}^2+\hat{W}'=W^2+W'$ gives rise to such an equation
through (\ref{hamiltonian SQM}). Nevertheless, the equation (\ref{vector
DeWolfe}) is uniquely specified by a requirement that the
effective potential $V_1$ is singular at $t=0$. This is true because
$V_1$ is singular in the KS case (\ref{vector potential}) and
hence should be singular everywhere on the branch by
continuity.}

We have calculated the spectrum of both equations numerically for
the backgrounds along the baryonic branch. Since the equations for
the superpartners are significantly different the discrepancy
between the masses can be used as an error estimate of the
numerical method used in the calculation.

\section{Numerical Analysis}
\label{numerics}
In this section we present the results  of the
numerical studies of bound state spectra for the baryonic branch
backgrounds. In our computations we will rely on the shooting
technique. The spectrum of the minimal scalar equation
(\ref{minimal scalar KS}) in the KS background was also studied
numerically in \cite{DM1,Krasnitz0,BHM2} while the analytical
approximation was employed in \cite{Tye}.

We start by comparing the KS spectra of the equations for graviton
(\ref{minimal scalar KS}) and vector mode (\ref{vector ks}). Two
fluctuations are related by supersymmetry and thus their masses
should be the same. The spectrum is presented in the table \ref{spectra table}. The eigenvalues
match with those obtained by Krasnitz \cite{Krasnitz0} with the WKB approximation. Comparing the numeric values of the
masses of the spin-2 and vector
particles in the table \ref{spectra table} one could estimate the error of the shooting technique
in the KS case to be around 0.1\%.

\begin{table}[h!b!p!t]
\begin{center}
\begin{tabular}{||c||c|c|c|c|c|c|c|c|c||}
  \hline
  % after \\: \hline or \cline{col1-col2} \cline{col3-col4} ...
  n & 1\Trule \Brule & 2 & 3 & 4 & 5 & 6 & 7 & 8 & 9
  %\hline
  \\ \hline Graviton \Trule \Brule& 1.764 & 4.002 & 7.143 & 11.19 & 16.16 & 22.03 & 28.83 & 36.54 & 45.16
  %\hline
  \\ Vector Mode \Trule \Brule  & 1.762 & 3.999 & 7.136 & 11.18 & 16.12 & 22.01 & 28.80 & 36.50 & 45.12
  \\ \hline
\end{tabular}
\caption{The spectrum of $m^2$ for the gravity multiplet}
\label{spectra table}
\end{center}
\end{table}
\vspace{-0.5cm}

\begin{figure}[htb]
\begin{center}
$\begin{array}{@{\hspace{-.3in}}c@{\hspace{-.3in}}c}
\epsfxsize=3.5in \epsffile{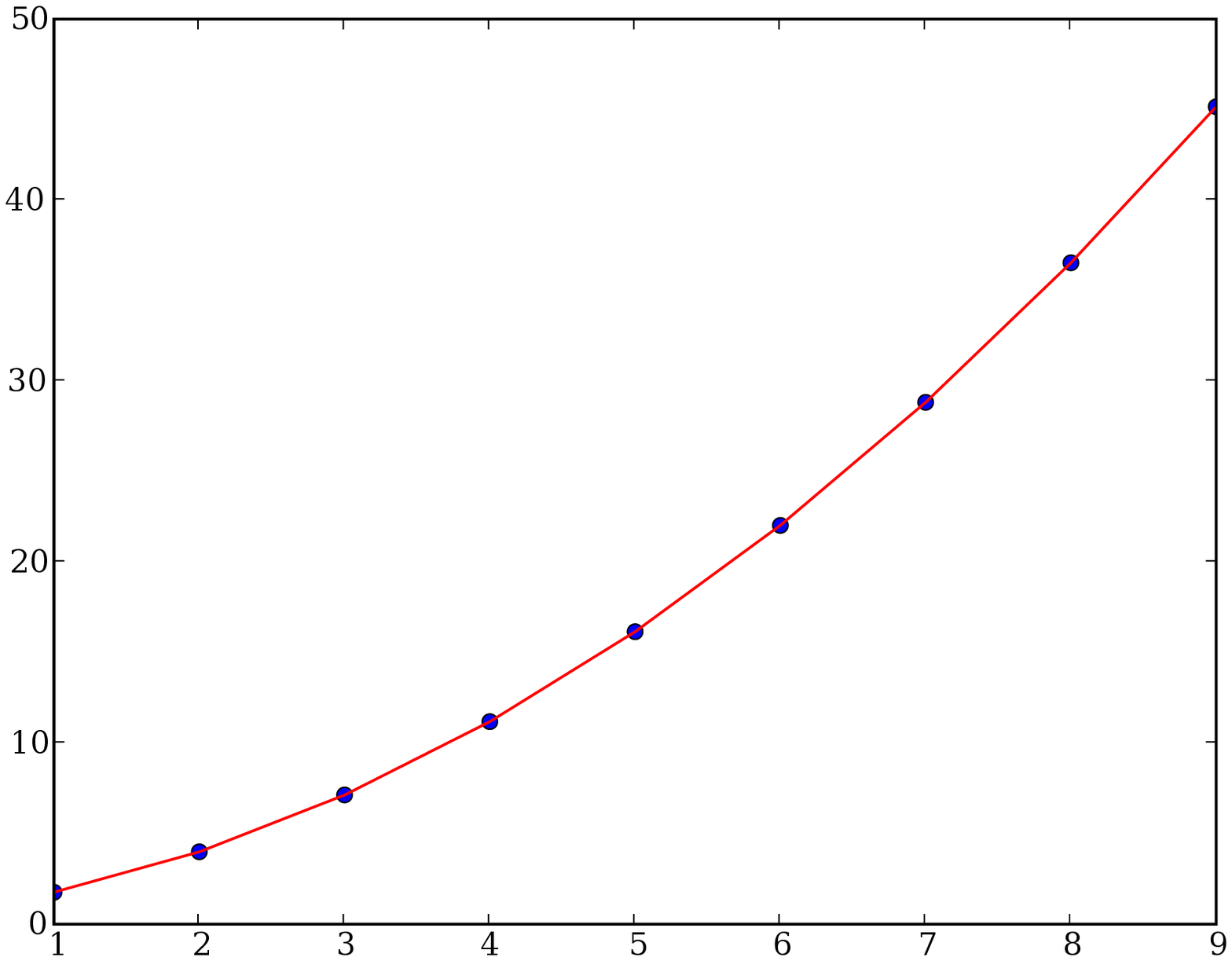} &
    \epsfxsize=3.0in
    \epsffile{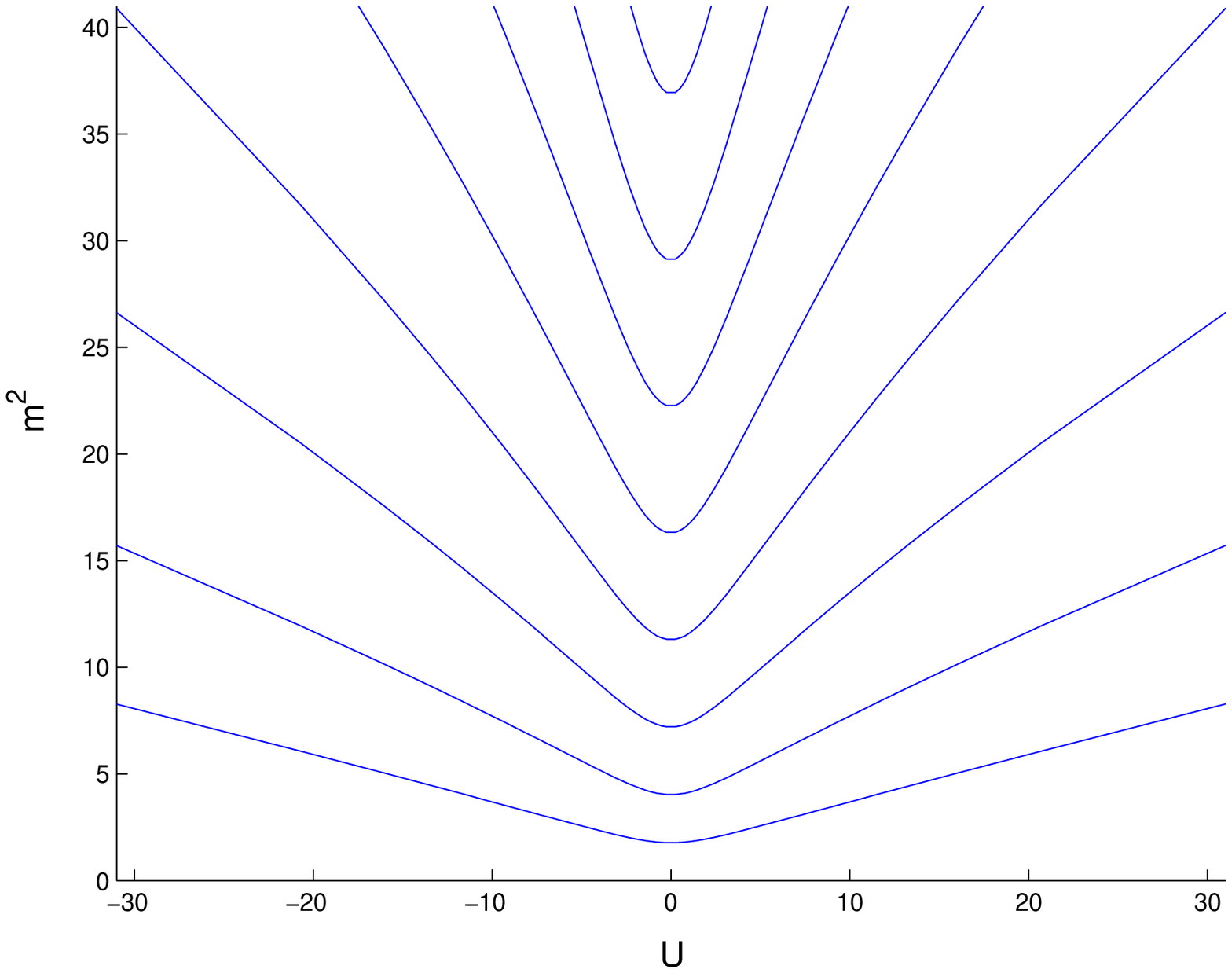} \\ [0.0cm]
\mbox{\footnotesize\bf (a)} & \mbox{\footnotesize\bf (b)}
\end{array}$
\end{center}
\vspace{-0.6cm}
\caption{\footnotesize{\bf(a)} Values of $m^2$ for the graviton multiplet in KS for different quantum numbers $n$. {\bf (b)}~Extension of the spectrum to the baryonic branch parameterized by $U$. }
\label{fig bb}
\end{figure}

First few (up to ten) values of $m^2$ in the KS spectrum can be approximated
with a good accuracy by a quadratic fit
\be
\label{quadratic fit}
m_n^2 = 0.46\,n^2 + 0.86\,n + 0.46, \qquad n=1,2,3,\ldots
\ee
We present the results of the fit and the masses on the figure
\ref{fig bb}(a). It is interesting that the fit (\ref{quadratic
fit}) is close to the spectrum even for small $n$. The fitting
formula (\ref{quadratic fit}) is proportional to $(n+n_0)^2$,
where $n_0$ is close to one. This is consistent with the
approximation of \cite{Tye}, where the eigenvalues were matched to
zeroes of the Bessel functions, ubiquitous in the conical geometry. A similar result was obtained
in \cite{FDW} for the GPPZ \cite{GPPZ} flow, where the exact
spectrum was proportional to $(n+1)^2$.

The fit (\ref{quadratic fit}) was found by minimizing the sum
\be \sum_{n=1}^N \left|m_n^2-(c_2n^2+c_1n+c_0)\right|^2
\ee
for the few first states $N=5,\ldots,10$. With more points taken into
account the least square fit would increase the accuracy of the
highest coefficient $c_2$ by the price of a larger deviation from
$m_n^2$ for small $n$.  We found $c_2$ to be $\sim 0.459$ in the
KS case. This number is in good agreement with the universal
coefficient obtained by Berg, Haack and M\"uck in \cite{BHM2}. In
their normalization the coefficient takes value
$(3/4)^{2/3}h(0)\,c_2\simeq 0.27$.

Remarkably, the coefficient $c_2$ does not depend on the details of
the effective potential, but rather encodes information about
the background geometry, namely, the combination $g^{00}g_{tt}$,
which arises from the Laplace operator in five dimensions.
Indeed, the WKB approach, applied in \cite{Krasnitz0}, gives
\be
\label{WKBint} \int_0^{t^*} \d t \left.
\sqrt{-V_2(t)}\,\right|_{\,k^2=-m^2_{n}}=\frac34\,\pi+(n-1)\pi,
\ee
where $V_2(t^*)=0$. In the KS case $V_2$ is given by
(\ref{graviton potential}). For large $n$, and consequently large
$m_n$, the $k^2$-independent term in $V_2$ can be dropped and we
obtain an analytical expression for $c_2$ in the KS case
\be
c_2=\pi^2 \left[\int_0^{\infty} \d t \,\frac{\sqrt{h} \sinh t}{ (\sinh
2t-2t)^{1/3}}\right]^{-2} \sim 0.460\ .
\ee

\begin{figure}[htb]
\begin{center}
$\begin{array}{c@{\hspace{.2in}}c}
\epsfxsize=3.05in
\epsffile{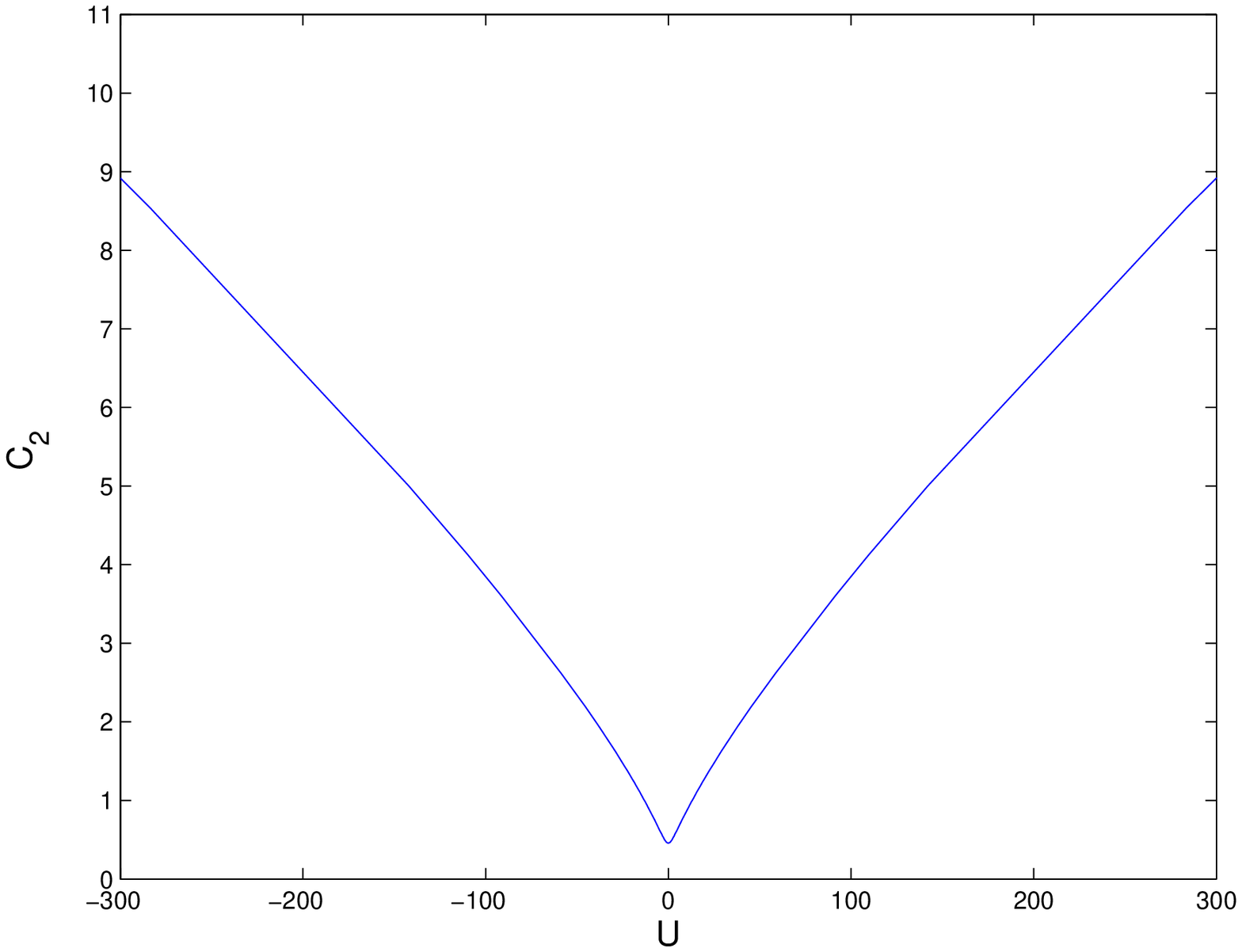}  &
    \epsfxsize=3.0in
    \epsffile{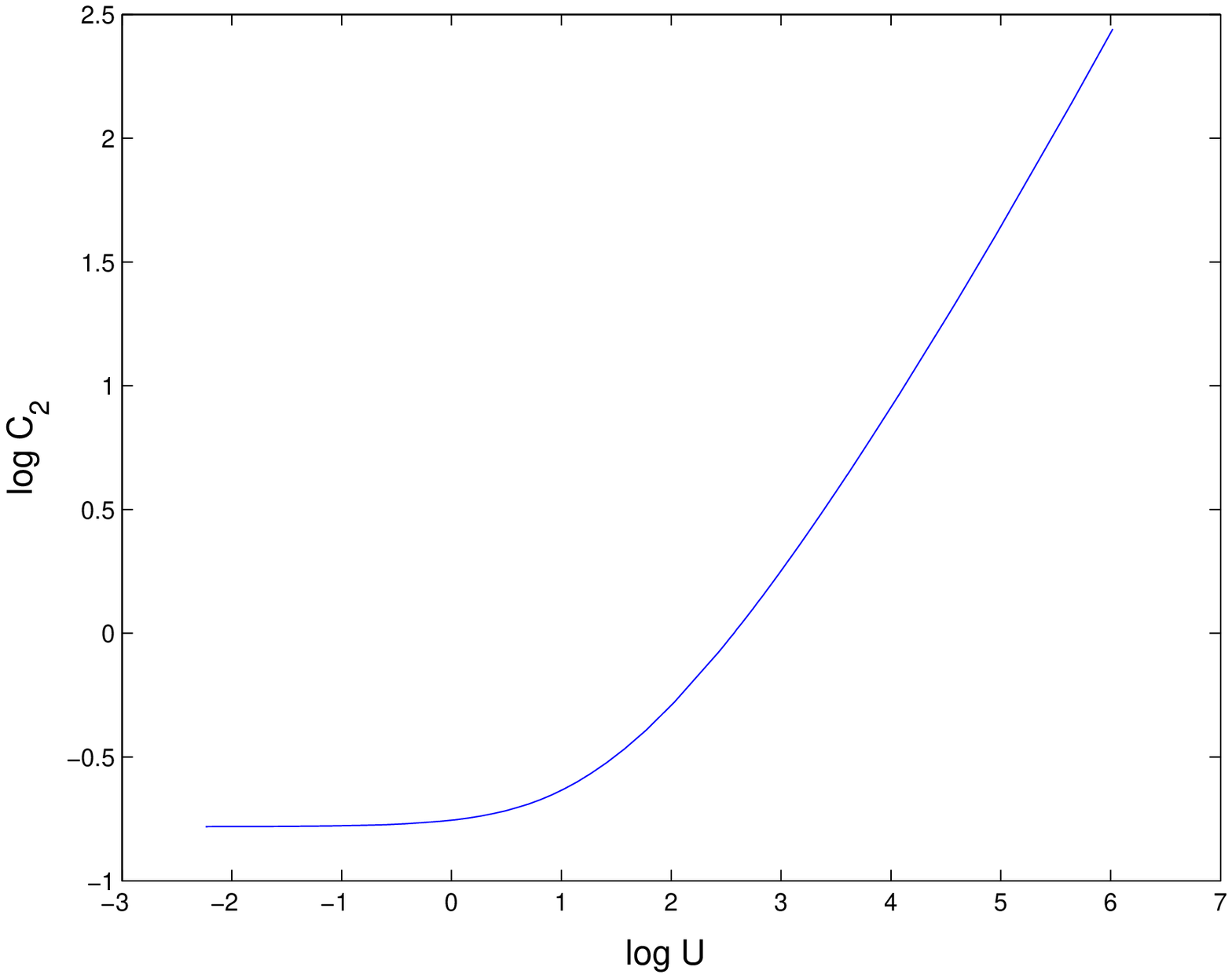} \\ [0.0cm]
\mbox{\footnotesize \bf (a)} & \mbox{\footnotesize \bf (b)}
\end{array}$
\end{center}
\vspace{-0.7cm}
\caption{\footnotesize {\bf(a)} $c_2$-coefficient as the function of $U$-parameter. {\bf (b)} $\log c_2$ as the function of $\log U$.}
\label{a1}
\end{figure}

Let us choose the coordinate $U$, introduced in \cite{DKS}, to
parameterize the baryonic branch. To estimate the scale of the
spectrum for a non-KS background we rewrite the potential
(\ref{graviton potential}) in terms of the  PT ansatz \cite{PT},
substituting $k^2$ for its eigenvalue $-m^2$:
\be
V_2(m^2,t)=-\frac{m^2e^{-2A+x}}{ v} +\frac{2a\cosh t}{
v}\,e^{-3g}-\frac{(a\cosh t+1)^2+2a^2\sinh^2t}{ v^2}\,e^{-2g},
\ee
where $a(t)$ is another function from the PT ansatz \cite{PT},
$e^{2g}=-1-a^2-2a\cosh t$, and $v=e^{6p+2x}$.

Although we cannot find the spectrum of $m^2$ analytically, we
can estimate how it scales with the parameter $U$ when we are
significantly far from the origin of the branch. We start our
analysis with the $m^2$-independent part of $V_2$, which only
slightly varies as we increase $U$. Indeed, its leading UV
($t\rightarrow\infty$) asymptotic is $U$-independent:
\be
V_2(0,t)=\frac49-\frac{(5-2t)}{ 6}\, U^2e^{-4t/3}+\ldots ;
\ee
and $V_2(0,t)$ varies within a small range in the IR ($t=0$):
\be
V_2(0,t)=\frac{1}{ 4}-\frac{3}{5}\,\xi(1-\xi)+\mathcal{O}(t^2).%+(\xi-1/6)(5/6-\xi)(9/35-108/175\xi(1-\xi))t^2
\ee
Here we remind that $\xi(U)\in (1/6\ldots 5/6)$ \cite{Butti}
is a function of $U$, which can also be used to parameterize the
branch. It varies within the specified limits, and the point
$\xi=1/2$ corresponds to the KS solution. Hence $V_2(0,0)=2/5$ for
KS and $V_2(0,0)$ approaches $1/3$ for large $U$. The
$V_2(0,t)$ is monotonic and therefore it can be approximated by a
constant in the analysis below.

Unlike $V_2(0,t)$, the mass-dependent component
$m^2\,e^{-2A+x}v^{-1}$ significantly depends on $U$. It
monotonically changes from a finite value at zero to the zero value
at infinity\footnote{Here we use the normalization of the warped
factor introduced by Krasntitz \cite{Krasnitz0}, what results in
$e^{-2A}= 2^{1/3}3 \sqrt{(e^{-2\Phi}-1)U^{-1}}$.}
\be
\label{u2 expansion uv}
e^{-2A+x}v^{-1}=\frac{2^{1/3}3}{ 16}\,
(4t-1)e^{-2t/3}+...
\ee

In general, the value at zero is a complicated function of $U$,
$\xi(U)$ and $\Phi_0=\Phi(U,t=0)$. It can be simplified in the
large $U$ range by substituting the limiting value $\xi=5/6$ and
expressing $\Phi_0$ in terms of $U$ and $\xi$ \cite{DKS}:
$e^{\Phi_0}\simeq 2^{3/2}3^{-1/4}U^{-3/4}$. This gives
\be
\label{u2 expansion}
e^{-2A+x}v^{-1}= \frac{2^{1/3}3}{
2U}\left[1-e^{2\Phi_0}\left(1+\frac{2t^2}{9}+\frac{2t^4}{
135}+...\right)\right].
\ee

The normalized  solution to the equation (\ref{Shroedinger form})
exist only if $V_2<0$ at the origin, which suggests that $m^2$
scales at least as $U$ for large $U$. We can try to be more
precise  using the semiclassical approximation and express the
$n$-th mass through the integral over the $V_2<0$ region, as we did
above in (\ref{WKBint}). This integral can be roughly approximated
as $\sqrt{-V_2(0)}t^* \sim {m t^*U^{-1/2}}$. The main complication
is to estimate $t^*$. Since $e^{2\Phi_0}$ from (\ref{u2
expansion}) is small, the perturbative expansion (\ref{u2
expansion}) suggests that $t^*$ increases with $U$ until a point,
where (\ref{u2 expansion}) is no longer reliable. At the same time
the large $t$ asymptotic (\ref{u2 expansion uv}) is
$U$-independent, what suggests that for large $U$ the value of
$t^*$ approaches a constant. Therefore we expect $m_{n}^2\sim
Un^2$ for sufficiently large $U$.

Numerical studies of the graviton multiplet spectrum on the
baryonic branch shows the pattern depicted in the figure \ref{fig
bb}(b). Calculations confirm that the leading coefficient $c_2$
grows as $U^{\alpha}$, where $\alpha$ approaches $1$ for large $U$
(figure \ref{a1}). As a final touch, we collect in the table
\ref{alpha table} the known evidence about the $U$ scaling
parameter $\alpha$ for some non-perturbative objects on the
baryonic branch.

\begin{table}[h!b!p!t]
\begin{center}
\begin{tabular}{cccccc}
  \hline\hline
      \Trule \Brule & SUSY $D5$ & Baryonic Condensate & Fundamental String & Glueballs & $D3,\bar{D}3$
  \\ \hline
  $\alpha$  \Trule \Brule & 0 & $\alpha<1$  & 1/4 & 1/2 & 5/4
  \\ \hline\hline
\end{tabular}
\caption{Scale behavior for large $U$: $T\sim U^\alpha$}
\label{alpha table}
\end{center}
\end{table}
\vspace{-0.5cm}

\section{Discussion}
\label{conclusions} In this work we present the equations
describing the bosonic degrees of freedom of the gravity multiplet
for the KS and the baryonic branch backgrounds. The equations were
derived by a linearization of the ten-dimensional type IIB
supergravity equations. The traceless graviton from the gravity
multiplet satisfies the equation for a scalar minimally coupled
to the background (\ref{minimal scalar PT}). The vector mode of
the gravity multiplet dual to the $\uor$ current satisfies the
equation (\ref{vector ks}) in the KS background. Its
generalization to the baryonic branch (\ref{eqn2 DeWolfe}) is
found by matching it to the equation (\ref{vector DeWolfe})
derived in the five-dimensional approach in \cite{DeWolfe}. This
result is supported by the supersymmetry transformation that
relates the wave functions of these fluctuations.

The mass spectrum of the gravity multiplet for the KS background
can be found in the table \ref{spectra table}. This spectrum can
be approximated with a good accuracy by a simple quadratic formula
(\ref{quadratic fit}), which is approximately
\be m_n^2\simeq
0.46(n+1)^2.
\label{full square}
\ee
This simple complete square form
does not hold along the baryonic branch, although the spectrum can be
well approximated by a general quadratic formula $c_2n^2+c_1n+c_0$.

In this work we did not study the spectrum of the anomaly
mutliplet $S$  of (\ref{superfield S}), which contains the
fluctuation of the trace of the metric $h_\mu^{~\mu}$ and the
longitudinal part of the vector fluctuation $\tilde{a}$ (\ref{vector
fluctuation}). The main complication is that these fluctuations do
not decouple from the other supergravity fields. Recently the
fluctuation of the metric $h_\mu^{~\mu}$ was considered as a part
of 7-particle system by Berg, Haack and M\"uck in
\cite{BHM1,BHM2}. They found the resulting spectrum of the system,
but the individual mass towers were not identified with the
glueballs.

Based on the similarities between the spectrum of the gravity
multiplet in the KS (\ref{full square}) and GPPZ backgrounds, where
$m_n^2=4L^{-2}(n+1)^2$, one can assume that some features of the
spectra for certain glueballs do not crucially depend on the
details of the background. Based on the exact result of the GPPZ
case calculation for the mass spectrum of the anomaly multiplet
$S$, $m_n^2=4L^{-2}(n+1)(n+2)$ \cite{DeWolfe,FDW,AFT} , one can guess
the answer for the KS case. In the units of BHM the approximate
formula reads
$$m_n^2\simeq 0.27(n+1)(n+2),\qquad n=1,2,3,\ldots$$
This is in fact close to the lightest of the seven towers of BHM, given by the empirical formula
$$0.271n^2+0.774n+0.562.$$
It would be interesting to confirm the matching between the trace
of the metric and the lowest tower of the 7-particle system with a
more rigorous approach.

We find it intriguing that the spectrum of \cite{BHM2} contains only
two states that look degenerate. Given that we are dealing with
massive states of the $\Nn=1$ system we would expect all the
states to be degenerate. This might signify that the numerical
method used in \cite{BHM2} alter the mass degeneracy because of a
numerical error. Another possibility is that the superpartners of
the glueballs in question are not the part of the 7-particle system
and can not be captured by the fluctuations of the PT ansatz. The
clarification of the magnitude of the numerical error is also
important to check another finding of \cite{BHM2} -- the
significant deviation of the spectrum from the quadratic behavior
for few lowest values of the quantum number $n$.

\vskip0.8cm
We are grateful to M.~Berg, O.~DeWolfe, M.~R.~Douglas,
D.~Z.~Freedman, I.~R.~Klebanov, A.~Konechny, G.~Moore and
N.~Seiberg for very useful comments and stimulating discussions.
The work was partly supported by Russian Federal Agency of Atomic
Energy; by the Council of the President of the Russian Federation
for Support of Leading Scientific Schools NSh-8004.2006.2, and by
the grants NSF PHY-0243680, RFBR 07-02-00878 (A.D.) and DOE
DE-FG02-96ER40949, RFBR 07-02-01161 (D.M.).


\newpage

\begin{thebibliography}{99}

\bibitem{Maldacena} J.~M.~Maldacena,
%``The large N limit of superconformal field theories and supergravity,''
Adv.\ Theor.\ Math.\ Phys.\  {\bf 2} (1998) 231 [Int.\ J.\ Theor.\
Phys.\  {\bf 38} (1999) 1113] [arXiv:hep-th/9711200]; S.~S.~Gubser, I.~R.~Klebanov and A.~M.~Polyakov,
  %``Gauge theory correlators from non-critical string theory,''
  Phys.\ Lett.\  B {\bf 428} (1998) 105
  [arXiv:hep-th/9802109]; E.~Witten,
  %``Anti-de Sitter space and holography,''
  Adv.\ Theor.\ Math.\ Phys.\  {\bf 2}, 253 (1998)
  [arXiv:hep-th/9802150].

\bibitem{KS} I.~R.~Klebanov and M.~J.~Strassler,
  %``Supergravity and a confining gauge theory: Duality cascades and
  %chiSB-resolution of naked singularities,''
  JHEP {\bf 0008} (2000) 052
  [arXiv:hep-th/0007191].

\bibitem{GHK} S.~S.~Gubser, C.~P.~Herzog and I.~R.~Klebanov,
  %``Symmetry breaking and axionic strings in the warped deformed conifold,''
  JHEP {\bf 0409}, 036 (2004)
  [arXiv:hep-th/0405282];   S.~S.~Gubser, C.~P.~Herzog and I.~R.~Klebanov,
  %``Variations on the warped deformed conifold,''
  Comptes Rendus Physique {\bf 5}, 1031 (2004)
  [arXiv:hep-th/0409186].


\bibitem{Butti} A.~Butti, M.~Grana, R.~Minasian, M.~Petrini and A.~Zaffaroni,
  %``The baryonic branch of Klebanov-Strassler solution: A supersymmetric
  %family of SU(3) structure backgrounds,''
  JHEP {\bf 0503}, 069 (2005)
  [arXiv:hep-th/0412187].

\bibitem{DKS} A.~Dymarsky, I.~R.~Klebanov and N.~Seiberg,
  %``On the moduli space of the cascading SU(M+p) x SU(p) gauge theory,''
  JHEP {\bf 0601}, 155 (2006)
  [arXiv:hep-th/0511254].

\bibitem{FZ} S.~Ferrara and B.~Zumino,
  %``Transformation Properties Of The Supercurrent,''
  Nucl.\ Phys.\  B {\bf 87} (1975) 207.


\bibitem{BDK}
  M.~K.~Benna, A.~Dymarsky and I.~R.~Klebanov,
  %``Baryonic condensates on the conifold,''
  arXiv:hep-th/0612136.


\bibitem{PT} G.~Papadopoulos and A.~A.~Tseytlin,
  %``Complex geometry of conifolds and 5-brane wrapped on 2-sphere,''
  Class.\ Quant.\ Grav.\  {\bf 18}, 1333 (2001)
  [arXiv:hep-th/0012034].

\bibitem{MN} A.~H.~Chamseddine and M.~S.~Volkov,
  %``Non-Abelian BPS monopoles in N = 4 gauged supergravity,''
  Phys.\ Rev.\ Lett.\  {\bf 79} (1997) 3343
  [arXiv:hep-th/9707176]; A.~H.~Chamseddine and M.~S.~Volkov,
  %``Non-Abelian solitons in N = 4 gauged supergravity and leading order  string
  %theory,''
  Phys.\ Rev.\ D {\bf 57} (1998) 6242
  [arXiv:hep-th/9711181]; J.~M.~Maldacena and C.~Nunez,
  %``Towards the large N limit of pure N = 1 super Yang Mills,''
  Phys.\ Rev.\ Lett.\  {\bf 86} (2001) 588
  [arXiv:hep-th/0008001].


\bibitem{DeWolfe} M.~Bianchi, O.~DeWolfe, D.~Z.~Freedman and K.~Pilch,
  %``Anatomy of two holographic renormalization group flows,''
 {JHEP}  {\bf 0101}, 021 (2001)
 [arXiv:hep-th/0009156].

\bibitem{Ouyang} I.~R.~Klebanov, P.~Ouyang and E.~Witten,
  %``A gravity dual of the chiral anomaly,''
 {Phys.\ Rev.\ D} {\bf 65}, 105007 (2002) [arXiv:hep-th/0202056].

\bibitem{GKT} S.~S.~Gubser, I.~R.~Klebanov and A.~A.~Tseytlin,
  %``String theory and classical absorption by three-branes,''
  Nucl.\ Phys.\ B {\bf 499}, 217 (1997)
  [arXiv:hep-th/9703040]; R.~C.~Brower, S.~D.~Mathur and C.~I.~Tan,
  %``Discrete spectrum of the graviton in the AdS(5) black hole background,''
  Nucl.\ Phys.\ B {\bf 574}, 219 (2000)
  [arXiv:hep-th/9908196]; N.~R.~Constable and R.~C.~Myers,
  %``Spin-two glueballs, positive energy theorems and the AdS/CFT
  %correspondence,''
  JHEP {\bf 9910}, 037 (1999)
  [arXiv:hep-th/9908175]; O.~DeWolfe, D.~Z.~Freedman, S.~S.~Gubser and A.~Karch,
  %``Modeling the fifth dimension with scalars and gravity,''
  Phys.\ Rev.\  D {\bf 62} (2000) 046008
  [arXiv:hep-th/9909134].

\bibitem{Tye} H.~Firouzjahi and S.~H.~Tye,
  %``The shape of gravity in a warped deformed conifold,''
  JHEP {\bf 0601}, 136 (2006)
  [arXiv:hep-th/0512076].

\bibitem{Krasnitz thesis} M.~Krasnitz,
  %``Correlation functions in a cascading N = 1 gauge theory from
  %supergravity,''
  JHEP {\bf 0212} (2002) 048
  [arXiv:hep-th/0209163].


\bibitem{BHM1} M.~Berg, M.~Haack and W.~Muck,
  %``Bulk dynamics in confining gauge theories,''
  Nucl.\ Phys.\  B {\bf 736}, 82 (2006)
  [arXiv:hep-th/0507285].


\bibitem{DM1} A.~Dymarsky and D.~Melnikov,
  %``On the glueball Spectrum in the Klebanov-Strassler Model,''
  preprint ITEP-TH-113/05, JETP\ Lett.\ {\bf 84}, 368 (2006).

\bibitem{Krasnitz0} M.~Krasnitz,
  %``A two point function in a cascading N = 1 gauge theory from
  %supergravity,''
  arXiv:hep-th/0011179.

\bibitem{Ceresole}
  A.~Ceresole and G.~Dall'Agata,
  %``General matter coupled N = 2, D = 5 gauged supergravity,''
  Nucl.\ Phys.\  B {\bf 585} (2000) 143
  [arXiv:hep-th/0004111].

\bibitem{BHM2} M.~Berg, M.~Haack and W.~Muck,
  %``Glueballs vs. gluinoballs: Fluctuation spectra in non-AdS/non-CFT,''
  arXiv:hep-th/0612224.



\bibitem{FDW} O.~DeWolfe and D.~Z.~Freedman,
    %``Notes on fluctuations and correlation functions in holographic
  %renormalization group flows,''
  arXiv:hep-th/0002226.


\bibitem{GPPZ} L.~Girardello, M.~Petrini, M.~Porrati and A.~Zaffaroni,
  %``The supergravity dual of N = 1 super Yang-Mills theory,''
  Nucl.\ Phys.\  B {\bf 569} (2000) 451
  [arXiv:hep-th/9909047].

\bibitem{AFT} G.~Arutyunov, S.~Frolov and S.~Theisen,
  %``A note on gravity-scalar fluctuations in holographic RG flow  geometries,''
  Phys.\ Lett.\ B {\bf 484}, 295 (2000)
  [arXiv:hep-th/0003116].






\end{thebibliography}
\end{document}